\documentclass[10pt]{elsarticle}

\usepackage{fullpage}
\usepackage{parskip}
\usepackage{amsmath}
\usepackage{amssymb}
\usepackage{enumerate}
\usepackage{graphicx}
\usepackage{psfrag}
\usepackage{caption}
\usepackage{subcaption}
\usepackage{color}
\usepackage{MnSymbol}
\usepackage{tikz}
\usepackage{bm}
\usepackage{multirow}
\usepackage{makecell}
\usepackage{array}
\usepackage{pgfplots}
  \pgfplotsset{compat=newest}
  \usetikzlibrary{plotmarks}
  \usepackage{grffile}
  \usepackage[titletoc,toc,title]{appendix}

\newcommand\solidrule[1][0.5cm]{\rule[0.5ex]{#1}{0.6pt}}

\newcommand{\mytriangle[1]}{\tikz{\draw[black,fill=#1] (0,0) -- (0.15cm,0) -- (0.075cm,0.15cm) -- cycle;}}
\newcommand{\mysquare[1]}{\tikz{\draw[black,fill=#1] (0,0) -- (0.13cm,0) -- (0.13cm,0.13cm) -- (0cm,0.13cm) -- cycle;}}

\biboptions{sort&compress}
\newlength\figH
\newlength\figW

\begin{document}

\begin{frontmatter}
\title{A strongly-coupled immersed-boundary formulation for thin deforming surfaces, with application to elastic beams}

\author[ad1]{Andres Goza\corref{cor1}}
\cortext[cor1]{Corresponding author}
\ead{ajgoza@gmail.com}
\author[ad1]{Tim Colonius}

\address[ad1]{Division of Engineering and Applied Science, California Institute of Technology, Pasadena, CA 91125, USA}

\begin{abstract}
We present a strongly-coupled immersed-boundary method for flow-structure interaction problems involving thin deforming bodies. The method is stable for arbitrary choices of solid-to-fluid mass ratios and for large body motions. As with many strongly-coupled immersed-boundary methods, our method requires the solution of a nonlinear algebraic system at each time step. The system is solved through iteration, where the iterates are obtained by linearizing the system and performing a block LU factorization. This restricts all iterations to small-dimensional subsystems that scale with the number of discretization points on the immersed surface, rather than on the entire flow domain. Moreover, the iteration procedure we propose does not involve heuristic regularization parameters, and has converged in a small number of iterations for all problems we have considered. We derive our method for general deforming surfaces, and verify the method with two-dimensional test problems of geometrically nonlinear beams undergoing large amplitude flapping behavior. 
\end{abstract}


\end{frontmatter}

\section{Introduction}

The immersed-boundary (IB) method is commonly used for flow-structure interaction (FSI) problems because of its ability to handle the fluid and immersed body separately. This flexibility comes at the cost of complicating the implementation of the no-slip boundary condition. This interface constraint nonlinearly couples the fluid and solid, and is therefore difficult to impose efficiently and robustly. We restrict our attention here to strongly-coupled IB methods, which strictly enforce the constraint. Weakly coupled IB methods do not impose the constraint, and are unstable for small solid-to-fluid mass ratios and large body motions \cite{causin, forster, borazjani}.

Due to the nonlinear nature of the constraint, most strongly-coupled methods must solve a large nonlinear system of equations at each time step. The block Gauss-Seidel procedure is one approach to solving this nonlinear system. It is attractive for its ease of implementation, but requires relaxation to converge for a wide range of mass ratios. Employing relaxation requires a heuristically chosen parameter, and can involve dozens of iterations to converge for small mass ratios \cite{tian14}, though Wang and Eldredge \cite{eldredge15} improved this convergence behavior using information about the system's added mass. Alternatively, the nonlinear system can be solved with a Newton-Raphson method. This removes the need of free parameters, and typically requires a small number of iterations irrespective of the mass ratio. However, this approach often involves computing several matrix-vector products per time step, each involving large Jacobian matrices \cite{degroote,mori,layton}. 

In the context of rigid body FSI problems, some strongly-coupled methods evaluate the constraint equation at the previous time step, which allows for the resulting equations of motion to be linear \cite{stern,taira16}. Thus, these methods do not require iterations, though the temporal accuracy is reduced to first order due to the time lag introduced in the constraint. It is difficult to avoid iteration in deforming body problems, since the solid equations have a nonlinear stiffness term that is frequently treated implicitly when discretized in time.

We present a strongly-coupled IB method for thin deforming surfaces that iteratively solves the nonlinear algebraic equations using a linearization of the system, as is done in the Newton-Raphson approach. Therefore, our method does not require free relaxation parameters, and has exhibited fast convergence behavior for all problems we have considered so far. A distinction between our method and a standard Newton-Raphson implementation is that we avoid large Jacobian matrices by performing a block LU factorization of the linearized system. This reduces all iterations to subsystems whose dimensions scale with the number of discretization points on the immersed surface, rather than on the entire flow domain. 

Our method treats the fluid with the two-dimensional (2D) discrete streamfunction formulation of Colonius and Taira \cite{tim08}, and the solid with a finite element formulation that applies to various solid materials undergoing  large deformations and rotations. We verify our method on 2D test problems of flow past deforming beams. The test problems involve large body motions and a wide range of mass ratios, and for all cases our proposed iteration process required a small number of iterations to converge. To supplement the 2D method presented in the main text, we derive in appendix \ref{app:app} an analogous method that treats the fluid with primitive variables. This method has a similar iteration procedure to the proposed 2D formulation, and can be applied in both 2D and 3D.

\section{Governing equations}
\label{sec:eqns}

We consider a fluid domain $\Omega$ and an immersed body $\Gamma$. We let $\textbf{x}$ denote the Eulerian coordinate representing a position in space, and define $\bm{\chi}(s,t)$ as the Lagrangian coordinate attached to the body $\Gamma$ ($s$ is a variable that parametrizes the surface). The dimensionless governing equations are written as
\begin{gather}
\frac{\partial \textbf{u}}{\partial t} + \textbf{u}\cdot \nabla \textbf{u} = - \nabla p + \frac{1}{\text{Re}} \nabla^2 \textbf{u} + \int_{\Gamma} \textbf{f}(\bm{\chi}(s,t)) \delta(\bm{\chi}(s,t) - \textbf{x}) ds  \label{eqn:NS} \\
\nabla \cdot \textbf{u} = 0 \label{eqn:incomp} \\ 
\frac{\rho_s}{\rho_f} \frac{\partial^2 \bm{\chi}}{\partial t^2} = \frac{1}{\rho_f U_\infty^2}\nabla \cdot \boldsymbol\sigma + \textbf{g}(\bm{\chi}) - \textbf{f}(\bm{\chi}) \label{eqn:sol_eqn} \\
\int_\Omega \textbf{u}(\textbf{x})\delta(\textbf{x} - \bm{\chi}(s,t)) d\textbf{x} = \frac{\partial \bm{\chi}(s,t)}{\partial t} \label{eqn:BC}
\end{gather}

In the above, $\textbf{x}$, $\bm{\chi}$, and $s$ were nondimensionalized by a characteristic length scale, $L$; \textbf{u} was nondimensionalized by a characteristic velocity scale, $U_\infty$; $t$ was nondimensionalized by $L/U_\infty$; and $p$ and \textbf{f} were nondimensionalized by $\rho_fU_\infty^2$, where $\rho_f$ is the fluid density ($\rho_s$ is the solid density).  The Reynolds number in (\ref{eqn:NS}) is defined as $\text{Re} = U_\infty L/ \nu$, where $\nu$ is the kinematic viscosity of the fluid. The term $\textbf{g}$ represents a body force per unit volume (e.g. gravity) applied to the immersed body, and was nondimensionalized by $\rho_f U_\infty^2/L$. The momentum equations (\ref{eqn:sol_eqn}) were divided by $\rho_f U_\infty^2/L$ to arrive at the form seen above.

In (\ref{eqn:sol_eqn}), the time derivative is understood to be a Lagrangian derivative, and the stress tensor used is the Cauchy stress, which is related to the second Piola-Kirchoff stress in the undeformed configuration, $\boldsymbol\sigma^K$, by
\begin{equation}
\sigma_{mn} = \frac{1}{J} \frac{\partial {x}_m}{\partial \chi^0_i}\frac{\partial x_n}{\partial \chi^0_j} \sigma_{ij}^K
\end{equation}
where $\bm{\chi}^0$ is the position of the body in its undeformed configuration, and $J = \det(\partial x_i/\partial \chi^0_j)$. The second Piola-Kirchoff stress is related to the strains within the solid via
\begin{equation}
\sigma_{ij}^K = D_{ijkl} E_{kl}
\end{equation}
where $\textbf{D}$ depends on Young's Modulus, $E$, the bulk shear modulus, $G$, and Poisson's ratio, $\nu_s$; and $\textbf{E}$ is the strain tensor given by
\begin{equation}
E_{ij} = \frac{1}{2}\left( \frac{\partial \chi_i}{\partial \chi^0_j} + \frac{\partial \chi_j}{\partial \chi^0_i} + \frac{\partial \chi_m}{\partial \chi^0_i}\frac{\partial \chi_m}{\partial \chi^0_j} \right)
\label{eqn:strain}
\end{equation} 
(summation implied on repeated indices). Again, all variables that comprise \textbf{E} were nondimensionalized using the characteristic length $L$.

The boundary condition on the interface $\Gamma$ is explicitly written as a constraint in (\ref{eqn:BC}). Its role is analogous to that of the continuity equation in computing the pressure: it is used to solve for the singular source term $\textbf{f}(\bm{\chi})$ that enforces the boundary condition on $\Gamma$. Note that $\textbf{f}(\bm{\chi})$ represents the surface stress imposed on the fluid by the immersed body. This can be seen by multiplying the momentum equations (\ref{eqn:NS}) by $\delta(\textbf{x} - \bm{\chi}(s,t))$ and integrating over the domain $\Omega$ to get
\begin{equation}
\textbf{f}(\bm{\chi}(s,t)) = \left[ \left( \frac{\partial}{\partial t} - \frac{1}{Re} \nabla^2 \right)\textbf{u} + \textbf{u}\cdot\nabla\textbf{u} + \nabla p     \right] \bigg|_{\textbf{x} = \bm{\chi}(s,t) } \qquad \forall \; \bm{\chi} \in \partial \Gamma
\label{eqn:fl_str}
\end{equation}
The negative of this term represents the stress imposed on the immersed body by the fluid, including added mass effects. The immersed-boundary method places fictitious fluid inside of bodies that contributes to the surface stress in (\ref{eqn:fl_str}). We restrict our attention to thin bodies, such as shells or membranes, where the fictitious fluid may be neglected in (\ref{eqn:fl_str}). Treating thick bodies would require the ability to remove the stress contribution from the fictitious fluid in (\ref{eqn:fl_str}) to obtain the stress from the physical fluid. This was done for rigid body FSI problems by L$\overline{\mbox{a}}$cis \emph{et al.}\ \cite{taira16}, but is difficult to extend to the deforming body case.

\section{Numerical method}
\label{sec:numerical}

In this section, we discretize the equations of motion in space to arrive at the coupled semi-discrete equations of motion: the fluid equations are discretized using the 2D discrete streamfunction formulation of Colonius and Taira \cite{tim08}, and the solid equations are discretized using a finite element formulation. We then discretize in time and introduce an efficient iteration procedure for solving the resulting nonlinear algebraic equations. 

\subsection{Semi-discrete equations}
\label{sec:semi_disc}

We begin by spatially discretizing the equations of motion for the fluid on a uniform Cartesian grid as
\begin{gather}
\dot{u} + \mathcal{N}(u) = -Gp + Lu + H(\chi) \tilde{f} \label{eqn:NSdisc_1} \\
Du = 0 \label{eqn:NSdisc_2} 
\end{gather} 
where the overdot denotes differentiation with respect to time, $u,p,\chi$ and $\tilde{f}$ denote the spatially discrete velocity, pressure, body position, and surface stresses; $\mathcal{N}(u)$ is a discretization of the nonlinear term; $G$, $L$, and $D$ are discretizations of the gradient, Laplacian, and divergence operators, respectively; and $H(\chi)$ is a discretization of the term involving the delta function in (\ref{eqn:NS}). The convective term is discretized in standard convective form with a central difference approximation of the derivatives, $L$ is built using the common 5 point finite difference stencil, and $G$ and $D$ are constructed with finite difference schemes such that $G = -D^T$. For computational efficiency, $H$ is built to satisfy $H\tilde{f} = - E^T f$, where $f$ is equal to $\tilde{f}$ to within a scaling factor and the matrix $E$ is a discretization of the term involving the delta function in (\ref{eqn:BC}). See reference \cite{tim07} for more details. 

Following Colonius and Taira \cite{tim08}, we avoid the incompressibility constraint by using a discrete curl operator that lies in the null space of the discrete divergence operator $D$. That is, $DC \equiv 0$, which implies that $C^TG = -(DC)^T \equiv 0$. The discrete curl operator engenders the use of a discrete streamfunction that is related to the discrete velocity field by $u = Cs$. Using this and premultiplying (\ref{eqn:NSdisc_1}) and (\ref{eqn:NSdisc_2}) by $C^T$ then gives the final form of the semi-discrete fluid equations that we consider:
\begin{gather}
C^TC\dot{s} + \mathcal{N}(Cs) = C^TLCs - C^TE^T(\chi)f \label{eqn:semi_disc_fluid1}
\end{gather}

The equations for the solid are discretized in space using a finite element procedure: the body is broken up into isoparametric finite elements with an associated set of compatible shape functions \cite{bathe}. A variable $(\cdot)$ belonging to the surface may be expressed with these basis functions as
\begin{equation}
(\cdot)(\bm{\chi}) = \sum_{j=1}^{N_{node}}  (\cdot)_j b_j(\bm{\chi})
\end{equation}
where $(\cdot)_j$ is the nodal value of $(\cdot)$ and $b_j$ is the shape function corresponding to node $j$. Using this expansion of each variable in (\ref{eqn:sol_eqn}), we arrive at a system of ordinary differential equations in time by multiplying (\ref{eqn:sol_eqn}) by the various shape functions and integrating over the volume of the immersed body. Letting $B$ be a matrix containing the various shape functions $b_j$, we write the spatially discretized form of (\ref{eqn:sol_eqn}) as
\begin{equation}
M\ddot{\chi}+ R(\chi) = Q(g + W(\chi)f)
\label{eqn:semi_disc_sol}
\end{equation}
where 
\begin{gather}
M = \frac{\rho_s}{\rho_f} \sum_{j=1}^{N_{el}} \int_{\Gamma^0_j}  B^TB d\bm{\chi}^0 \label{eqn:Mdef} \\
R(\chi) =  \frac{1}{\rho_f U_\infty^2} \sum_{j=1}^{N_{el}} \int_{\Gamma^0_j}  B_E^T \sigma^K d\bm{\chi}^0 \label{eqn:Rdef} \\
Q =   \sum_{j=1}^{N_{el}} \int_{\Gamma^0_j}  B^TB d\bm{\chi}^0  = \frac{\rho_f}{\rho_s} M\label{eqn:Qdef}
\end{gather}
In the above, $\Gamma^0_j$ denotes element $j$ of $\Gamma$ in its undeformed configuration, $B_E$ is a matrix containing the derivatives of the shape functions with respect to the nodal positions, and $\sigma^K$ contains nodal values of its continuous analog. Note that $\sigma^K$ is arranged as a vector so that $R(\chi)$ is also a vector. The nonlinearity of $R(\chi)$ is due to the dependence of $B_E$ and $\sigma_K$ on $\chi$. 

It is worth mentioning that the surface stresses $f$ obtained by many immersed-boundary methods contain spurious oscillations \cite{goza,seo,yang}. In the context of the present immersed-boundary method, these oscillations are due to the fact that the surface stresses are obtained from an ill-posed integral equation. We recently observed that for an appropriate choice of delta function, a weighting matrix, $W$, may be used to obtain accurate surface stresses \cite{goza}. Thus, $W$ is included in (\ref{eqn:semi_disc_sol}) to apply the physically correct surface stresses on the immersed surface. The specific form of $W$ is described in reference \cite{goza}, but we note here that $Wf$ is a discretization of 
\begin{equation}
\frac{\int_\Omega \int_\Gamma \textbf{f}(\bm{\chi}(s',t)) \delta_h(\textbf{x} - \bm{\chi}(s',t) ) \delta_h(\textbf{x} - \bm{\chi}(s,t)) ds' d\textbf{x} }{\int_\Gamma \delta_h(\textbf{x} - \bm{\chi}(s',t)) ds'}
\label{eqn:Wfdisc}
\end{equation}
where $\delta_h$ is a continuous delta function whose support depends on the grid spacing $h$ (see, for example, the review of Peskin \cite{peskin} for more information on continous delta functions and how to construct them). The expression in (\ref{eqn:Wfdisc}) is a spatially first order approximation to $\textbf{f}(\bm{\chi})$, so the presence of $W$ is a first order modification of (\ref{eqn:sol_eqn}). In the present work, we use the delta function of Roma \cite{roma}, which we found to provide a good combination of computational efficiency and accuracy of the surface stress.

Equation (\ref{eqn:semi_disc_sol}) and the corresponding definitions of $M$, $R$, and $Q$ are valid for a variety of solid materials undergoing large deformations, displacements, and rotations. In this work we restrict our attention to beams, for which we employ a corotational formulation \cite{crisfield}. In this formulation, arbitrarily large displacements and rotations are accommodated by attaching a local coordinate frame to each beam element. The strains are assumed to be small in this frame, and the corresponding internal stresses $R(\chi)$ are well known (see, \emph{e.g.}, reference \cite{crisfield}). Materials other than beams would require changes in the choice of elements, shape functions, and model for $\sigma^K$. However, these changes would not affect the structure of (\ref{eqn:semi_disc_sol}) or the ensuing time discretization procedure. 

Using these discretizations of the fluid and solid equations, the discrete form of the constraint (\ref{eqn:BC}) is written in terms of the streamfunction as
\begin{equation}
E(\chi)Cs - \zeta = 0 \label{eqn:BC_semidisc}
\end{equation}

Defining $\zeta:= \dot{\chi}$, the fully coupled FSI equations may be written as a first order system of differential-algebraic equations given by
\begin{gather}
C^TC\dot{s} = - \mathcal{N}(Cs) + C^TLCs - C^TE^T(\chi)f \label{eqn:eq_s} \\
M \dot{\zeta} = -R(\chi) + Q(g + W(\chi)f) \label{eqn:eq_u} \\
\dot{\chi} = \zeta \label{eqn:eq_x} \\
E(\chi)Cs - \zeta = 0 \label{eqn:eq_c} 
\end{gather}

Note that (\ref{eqn:NSdisc_1})--(\ref{eqn:NSdisc_2}) and (\ref{eqn:semi_disc_sol}) are valid in both 2D and 3D; the restriction to 2D is due only to the discrete streamfunction formulation (\ref{eqn:semi_disc_fluid1}). We derive in appendix \ref{app:app} an alternative method that treats the fluid using (\ref{eqn:NSdisc_1})--(\ref{eqn:NSdisc_2}). This method has an analogous iteration procedure to the one proposed in section \ref{sec:disc} and is readily extendible to 3D.

\subsection{Time discretization and efficient factorization procedure}
\label{sec:disc}

We discretize (\ref{eqn:eq_s}) using an Adams-Bashforth scheme for the nonlinear term and a Crank-Nicholson method for the diffusive term. Equations (\ref{eqn:eq_u})--(\ref{eqn:eq_x}) are discretized using an implicit Newmark scheme. The constraint equation (\ref{eqn:eq_c}) is evaluated at the current time step. That is, the method is strongly-coupled, which is necessary for the method to be stable for a wide range of mass ratios and in the presence of large body displacements \cite{causin, forster, borazjani}. We discretize the equations of motion in the aforementioned way to illustrate the iteration procedure on a commonly used scheme. However, the proposed iteration approach can readily be extended for a variety of time stepping schemes.

Discretizating (\ref{eqn:eq_s})--(\ref{eqn:eq_c}) as described in the previous paragraph leads to a system of nonlinear algebraic equations given by
\begin{gather}
C^TAC s_{n+1} + C^TE_{n+1}^T f_{n+1}= r^f_n \label{eqn:eq_sd} \\
\frac{4}{\Delta t^2} M \zeta_{n+1} + ( R(\chi_{n+1}) - QW_{n+1}f_{n+1} ) = r^\zeta_n \label{eqn:eq_ud} \\
\frac{2}{\Delta t} \chi_{n+1} - \zeta_{n+1} = r^\chi_n \label{eqn:eq_xd} \\
E_{n+1}Cs_{n+1} - \zeta_{n+1} = 0 \label{eqn:eq_cd} 
\end{gather}  
where $A = \frac{1}{\Delta t} I - \frac{1}{2}L$, $r^f_n = (\frac{1}{\Delta t} C^TC + \frac{1}{2} C^TLC) s_n + \frac{3}{2}C^T\mathcal{N}(Cs_n) - \frac{1}{2}C^T\mathcal{N}(Cs_{n-1})$, $r^\zeta_n = M(\frac{4}{\Delta t^2}\chi_n + \frac{4}{\Delta t} \zeta_n + \dot{\zeta}_n) + Qg $, and $r^\chi_n = \zeta_n + \frac{2}{\Delta t}\chi_n$. Note that the operators $E$, $E^T$, and $W$ are given subscripts to indicate their dependence on $\chi$.

We now describe an iteration procedure for using a guess for the solution at iteration $(k)$ to compute a new guess at iteration $(k+1)$ (We use the solution at time step $n$ as the guess for $(k) = 0$). To do this, we write $\chi_{n+1}^{(k+1)} = \chi_{n+1}^{(k)} + \Delta \chi$, $\zeta_{n+1}^{(k+1)}=\zeta_{n+1}^{(k)} + \Delta \zeta$, where the increments $\Delta \chi$, $\Delta \zeta$ are assumed to be small. Substituting this decomposition into (\ref{eqn:eq_sd})--(\ref{eqn:eq_cd}) and retaining first order terms in the increments and $\Delta t$ gives the linear system\footnote{Derivative terms that arise in the expansion of $E_{n+1}^{(k+1)}$, $E_{n+1}^{(k+1)T}$, and $W^{(k+1)}_{n+1}$ may be neglected to within $O(\Delta t)$. For example, the fourth block equation of (\ref{eqn:block_sys}) including this extra term is $E_{n+1}^{(k)}Cs_{n+1} + (DE_{n+1}^{(k)}\Delta \chi) Cs_{n+1} - \Delta \zeta = \zeta_{n+1}^{(k)}$. The $DE$ term is of order $O(\Delta t)$ compared with the $\Delta \zeta$ term, since by the third block equation $\Delta \chi = O(\Delta \zeta \Delta t)$.}
\begin{equation}
\renewcommand\arraystretch{1.6}
\begin{bmatrix} C^TAC & 0 & 0 & C^TE_{n+1}^{(k)T} \\ 0 & 0 & \frac{4}{\Delta t^2}M + K^{(k)} & -QW_{n+1}^{(k)} \\ 0 & -I & \frac{2}{\Delta t}I & 0 \\ E_{n+1}^{(k)}C & -I & 0 & 0 \end{bmatrix} \begin{bmatrix} s_{n+1} \\ \Delta \zeta \\ \Delta \chi \\ f_{n+1}^{(k+1)} \end{bmatrix} = \begin{bmatrix} r^f_n + O(\Delta t) \\ r^\zeta_n - \frac{4}{\Delta t^2} M\chi_{n+1}^{(k)} - R(\chi_{n+1}^{(k)} ) + O(\Delta t) \\ r_n^\chi - \frac{2}{\Delta t}\chi^{(k)}_{n+1} + \zeta_{n+1}^{(k)} \\ \zeta_{n+1}^{(k)} + O(\Delta t) \end{bmatrix} := \begin{bmatrix}  r^f_n \\ r^{\zeta\,(k)} \\ r^{\chi \, (k)} \\ r^{c \, (k)} \end{bmatrix} \label{eqn:block_sys}
\end{equation}
where $K^{(k)} = dR/d\chi|_{\chi = \chi_{n+1}^{(k)} }$. For beams this stiffness matrix has well known analytical expressions \cite{crisfield,bathe}.

The linear system (\ref{eqn:block_sys}) may be factored using a block LU decomposition. Defining $\hat{K}^{(k)} := \frac{4}{\Delta t^2} M+ K^{(k)}$ and $B_{n+1}^{(k)}:= E_{n+1}^{(k)}C(C^TAC)^{-1}C^TE_{n+1}^{(k)T}  $, the factored equations are
\begin{gather}
s^{*} = (C^TAC)^{-1}r^f_n \label{eqn:s*_fac} \\
\renewcommand\arraystretch{1.6}
\begin{bmatrix} B_{n+1}^{(k)} &  I \\ -\frac{2}{\Delta t} QW_{n+1}^{(k)} & \hat{K}^{(k)} \end{bmatrix} \begin{bmatrix} f_{n+1}^{(k+1)} \\ \Delta \zeta \end{bmatrix} = \begin{bmatrix} E_{n+1}^{(k)} Cs^* - r^{c\,(k)} \\ \frac{2}{\Delta t} r^{\zeta\,(k)} - r^{\chi\, (k)} \end{bmatrix} \label{eqn:fdu_fac} 
\renewcommand\arraystretch{1} \\
\Delta \chi = \frac{\Delta t}{2} (\Delta \zeta + r^{\chi \, (k)} ) \label{eqn:dx_fac} \\
s_{n+1} = s^{*} - (C^TAC)^{-1} C^TE_{n+1}^Tf_{n+1} \label{eqn:s_fac}
\end{gather}

Note that (\ref{eqn:s*_fac}) does not depend on information at time step $n+1$, and (\ref{eqn:fdu_fac})--(\ref{eqn:dx_fac}) do not require knowledge of $s_{n+1}$. Thus, (\ref{eqn:s*_fac}) can be solved once and for all at the beginning of each time step, and $s_{n+1}$ only needs to be computed once, after (\ref{eqn:fdu_fac})--(\ref{eqn:dx_fac}) have been iterated to convergence. This has the benefit that all iterations are restricted to (\ref{eqn:fdu_fac})--(\ref{eqn:dx_fac}), which have dimensions on the order of the number of body points, rather than the total number of points in the flow domain. We do not write a superscript on $s_{n+1}$ since it does not need to be iterated on. Wang \emph{et al.}\ \cite{eldredge15} also restricted iterations to small dimensional subsystems like (\ref{eqn:fdu_fac})--(\ref{eqn:dx_fac}). In their case, the solid equations were replaced by the rigid body equations of motion, and a block Gauss-Seidel procedure with added mass relaxation was used to solve their analogous nonlinear system.

A Poisson-like problem $(C^TAC)^{-1}$ must be solved in (\ref{eqn:s*_fac}), (\ref{eqn:s_fac}), and in each matrix-vector multiply with $B_{n+1}^{(k)}$. Solving the Poisson-like problem may be done efficiently using fast Fourier transforms, but requires operations of the order of the number of points on the flow domain. Thus, depending on the dimensions of the system, the computation and storage of $(\hat{K}^{(k)})^{-1}$ may be small compared with solving the Poisson-like problem. In this case one may perform block Gauss-elimination to reformulate (\ref{eqn:fdu_fac}) as
\begin{gather}
\left(B_{n+1}^{(k)} + \frac{2}{\Delta t}(\hat{K}^{(k)})^{-1}QW_{n+1}^{(k)}\right)f_{n+1}^{(k+1)} = E_{n+1}^{(k)}C s^{*} - r^{c\,(k)} -\frac{2}{\Delta t} (\hat{K}^{(k)})^{-1}  r^{\zeta \,(k)} + r^{\chi \,(k)}   \label{eqn:ffac_alt} \\
\Delta \zeta = \frac{2}{\Delta t} (\hat{K}^{(k)})^{-1} ( r^{\zeta \,(k)} + QW_{n+1}^{(k)}f_{n+1}^{(k)}) - r^{\chi \,(k)} \label{eqn:dufac_alt}
\end{gather}
This allows for  (\ref{eqn:ffac_alt}) and (\ref{eqn:dufac_alt}) to be solved sequentially. Moreover, while (\ref{eqn:ffac_alt}) must still be solve iteratively because of the embedded Poisson-like problem in $B_{n+1}^{(k)}$, the conditioning of the system is improved because the heterogeneous blocks of (\ref{eqn:fdu_fac}) are no longer present. 

We have tested both (\ref{eqn:fdu_fac}) and (\ref{eqn:ffac_alt})--(\ref{eqn:dufac_alt}), and found the formulation (\ref{eqn:ffac_alt})--(\ref{eqn:dufac_alt}) to be more efficient for the test problems considered. We solve (\ref{eqn:ffac_alt}) using the BiCGSTAB method, which typically requires 2--8 iterations to converge using $f_{n+1}^{(k)}$ as the initial guess. Moreover, the number of iterations required by BiCGSTAB often decreases as $k$ increases, since $f_{n+1}^{(k)}$ becomes an increasingly good guess for $f_{n+1}^{(k+1)}$.

It is also worth noting that the updates for $\chi_{n+1}^{(k)}$, $\zeta_{n+1}^{(k)}$, and $f_{n+1}^{(k)}$ do not depend on a heuristic relaxation parameter, as is often required when applying the block Gauss-Seidel iterative procedure to FSI solvers. Moreover, the proposed iteration procedure has required a small number of iterations to converge for all problems we have considered so far.

\section{Verification on flapping beam problems}

We verify our method for several test problems of 2D flow past deforming beams. The dynamics of this system are governed by the Reynolds number, dimensionless mass ratio, dimensionless bending stiffness, and Froude number. These are given, respectively, as
\begin{equation}
\text{Re} = \frac{U_\infty L}{\nu}, \; \text{M}_\rho = \frac{\rho_sh}{\rho_fL}, \; \text{K}_{\text{B}} = \frac{EI}{\rho_fU_\infty^2L^3}, \; \text{Fr} = \frac{U_\infty}{\sqrt{gL}}
\end{equation}
 where $h$ is the thickness of the beam, $EI$ is the bending stiffness of the beam, $\nu$ is the kinematic viscosity of the fluid, and $g$ is the gravitational constant.  
 
\begin{figure}[h!]
\centering
  \begin{subfigure}[b]{0.48\textwidth}
        \centering
        \includegraphics[height=60pt]{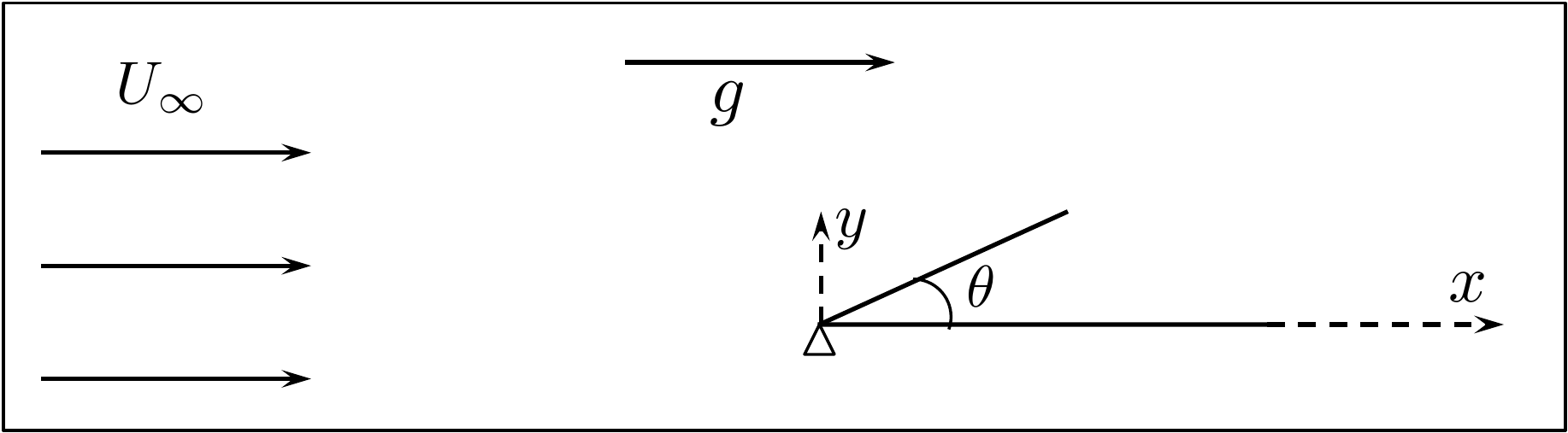}
    \end{subfigure}%
    \begin{subfigure}[b]{0.48\textwidth}
        \centering
        \includegraphics[height=60pt]{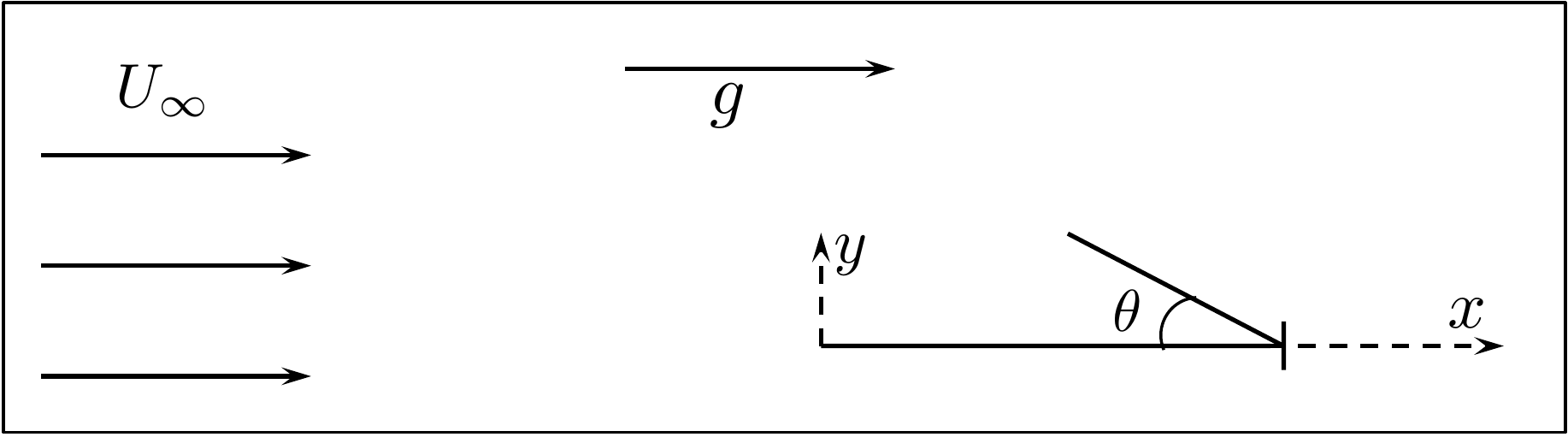}
    \end{subfigure}%
    \caption{Flow moves from left to right past a beam in the standard configuration (left) or the inverted configuration (right).}
    \label{fig:setup}
\end{figure}

We consider problems with the beam pinned at the leading edge (standard configuration) and with the beam clamped at the trailing edge (inverted configuration). Schematics of the different problem setups are shown in figure \ref{fig:setup}. 

Each configuration presents different challenges: beams in the standard configuration are typically associated with smaller mass ratios, whereas those in the inverted configuration often undergo larger motions and have greater dimensionless bending stiffnesses.   

The flow equations are solved using a multidomain approach: the finest grid surrounds the body and grids of increasing coarseness are used as distance from the body increases (see reference \cite{tim08} for details). For all cases, the immersed-boundary spacing is set to be twice that of the flow grid spacing on the finest sub-domain. We found that to give well-conditioned matrices $B_{n+1}^{(k)}$ without making the body porous (Kallemov \emph{et al.}\ \cite{griffith} also found this ratio to be preferable). We used a convergence criteria of $||\Delta \chi||_\infty/||\chi_{n+1}^{(k+1)}||_\infty \le 1\times10^{-7}$ when iterating between (\ref{eqn:dx_fac}), (\ref{eqn:ffac_alt}), and (\ref{eqn:dufac_alt}).

\subsection{Flow past a beam in the standard configuration}

We first consider flow past a deforming beam in the standard configuration with $\text{Re} = 1000$, $\text{M}_\rho = 0.075$, $\text{K}_\text{B} =0.0001$, and $\text{Fr} =0$ (\emph{i.e.}, no gravitational body force). This choice of parameters leads to limit cycle flapping, which we triggered in our simulations by initializing the beam at $\theta = 1^\circ$ (see figure \ref{fig:setup} for the definition of $\theta$). After an initial transient, the trailing edge of the beam has oscillatory transverse displacement of fixed amplitude and frequency. Table \ref{tab:Connell_LC} shows this amplitude and frequency as computed by various authors. Our finest domain was of size $[-0.2, 1.8] \times [-0.3, 0.3]$, and the entire flow domain size was $[-15.20,16.80]\times[-4.78,4.78]$. The grid spacing on the finest subdomain was $h=0.0025$, and the time step was $\Delta t = 0.0006$. Using a grid spacing of $h = 0.003$ changed our results in table \ref{tab:Connell_LC} by less than one percent. Five iterations of (\ref{eqn:dx_fac}), (\ref{eqn:ffac_alt}), and (\ref{eqn:dufac_alt}) were required for the first time step, and a maximum of three iterations were required for all remaining time steps.

\begin{table}[h!]
\begin{center}
\begin{tabular}{ c !{\vrule width1.2pt}c   c  }
 & Amplitude & Frequency ($St$) \\ \Xhline{3\arrayrulewidth}
Connell \emph{et al.}\ \cite{connell} & $\pm 0.096$ & $0.93$ \\ 
Gurugubelli \emph{et al.}\ \cite{gurugubelli} & $\pm 0.098$ & $0.95$ \\ 
Present & $\pm 0.097$ & $0.94$ \\
\end{tabular}
\caption{Amplitude and frequency associated with the transverse displacement of the beam's trailing edge; obtained for Re $ = 1000$, M$_\rho = 0.075$, \text{K}$_\text{B} =0.0001$, and Fr $ =0$.
}
\label{tab:Connell_LC} 
\end{center}
\end{table}

Limit cycle flapping also occurs for $\text{Re}= 200$, $\text{M}_\rho = 1.5$, $\text{K}_\text{B} =0.0015$, and $\text{Fr} = 1.4$. We show in table 2 the associated amplitude and frequency of the transverse trailing edge displacement. For comparison with the literature, we initialized the beam at $\theta = 18^\circ$. Our finest sub-domain for this problem was of size $[-0.2,1.8]\times[-0.9,0.9]$, and the total domain size was $[-15.20,16.80]\times[-7.95,7.95]$. The grid spacing on the finest domain was $h = 0.00625$, and the time step was $\Delta t = 0.0001$. Using $h = 0.005$ changed our reported results from table \ref{tab:Huang_LC} by less than one percent. Three iterations of (\ref{eqn:dx_fac}), (\ref{eqn:ffac_alt}), and (\ref{eqn:dufac_alt}) were required for the first time step, and a maximum of two iterations were required for all remaining time steps.

\begin{table}[h!]
\begin{center}
\begin{tabular}{ c !{\vrule width1.2pt}c   c }
 & Amplitude & Frequency ($St$) \\ \Xhline{3\arrayrulewidth}
Huang \emph{et al.}\ \cite{huang} & $\pm 0.35$ & $0.30$ \\ 
Wang \emph{et al.}\ \cite{eldredge15} & $\pm 0.35$ & $0.31$ \\ 
Lee \emph{et al.}\ \cite{lee} & $\pm 0.38$ & $0.31$ \\  
Present & $\pm 0.38$ & $0.32$ \\ 
\end{tabular}
\caption{Amplitude and frequency associated with the transverse displacement of the beam's trailing edge; obtained for Re $ = 200$, M$_\rho = 1.5$, \text{K}$_\text{B} =0.0015$, and Fr $ = 1.4$.
}
\label{tab:Huang_LC} 
\end{center}
\end{table}

\begin{figure}[h!]
\centering
\includegraphics[width=0.9\textwidth]{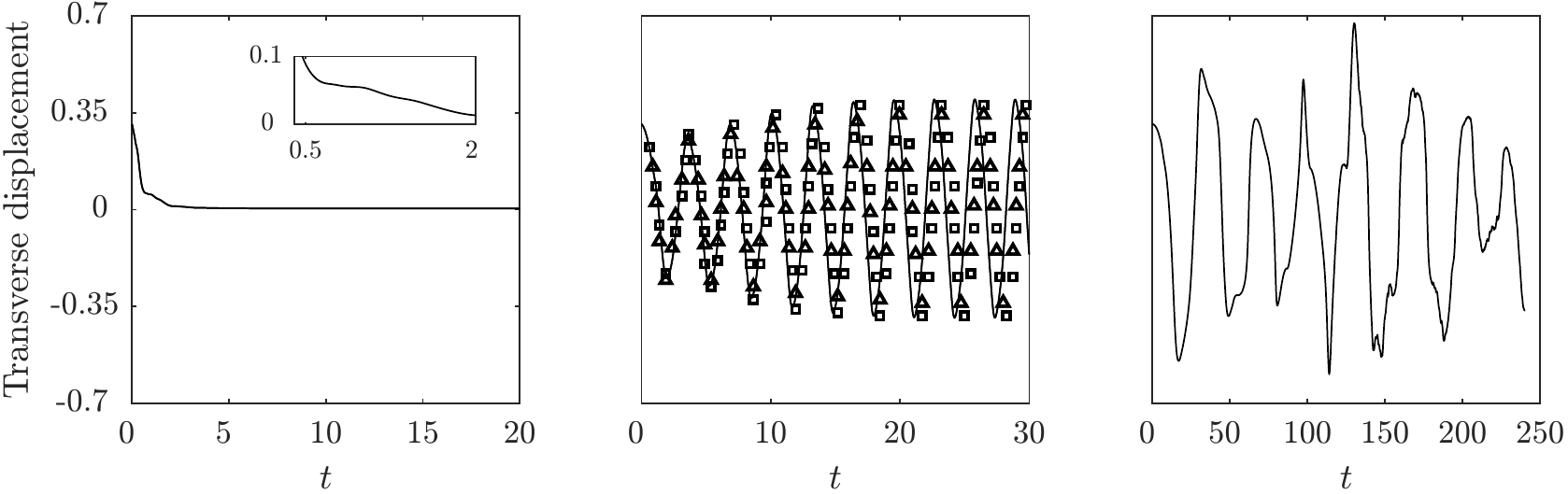}
\caption{Transverse displacement of the beam's trailing edge for Re$ = 200$, $\text{K}_\text{B} = 0.0015$, and $\text{Fr} = 1.4$. Left panel: $\text{M}_\rho = 0.0001$, middle panel: $\text{M}_\rho = 1.5$, right panel: $\text{M}_\rho = 100$. Present: \protect\solidrule[3mm], Lee \emph{et al.}\ \cite{lee}: \protect\mysquare[none], Huang \emph{et al.}\ \cite{huang}: \protect\mytriangle[none]. The insert on the left panel is a zoom-in of the $\text{M}_\rho = 0.0001$ case for $t\in [0.5, 2]$. Note the different horizontal axis values on each panels.  }
\label{fig:Huang_LC_fig}
\end{figure}

Figure \ref{fig:Huang_LC_fig} gives a time history of the trailing edge transverse displacement for the parameters corresponding to table \ref{tab:Huang_LC} (results from references \cite{lee,huang} are included for comparison). To emphasize the robustness of the method for a range of mass ratios, figure \ref{fig:Huang_LC_fig} also shows the trailing edge displacement for $\text{M}_\rho = 0.0001$ and $\text{M}_\rho = 100$. To our knowledge, the $\text{M}_\rho = 0.0001$ and $\text{M}_\rho = 100$ cases have not been simulated before. Moreover, we were not able to simulate the $\text{M}_\rho = 0.0001$ case when solving (\ref{eqn:eq_sd})--(\ref{eqn:eq_cd}) with the Gauss-Seidel method, even when using extensive relaxation. Using (\ref{eqn:dx_fac}), (\ref{eqn:ffac_alt}), and (\ref{eqn:dufac_alt}), a maximum of three iterations per time step were required after the first twenty time steps. During the first twenty time steps, up to fifteen iterations were required due to the impact of the impulsive start on a beam with such small inertia.

\begin{figure}[h!]
\centering
    \includegraphics[width=1\textwidth]{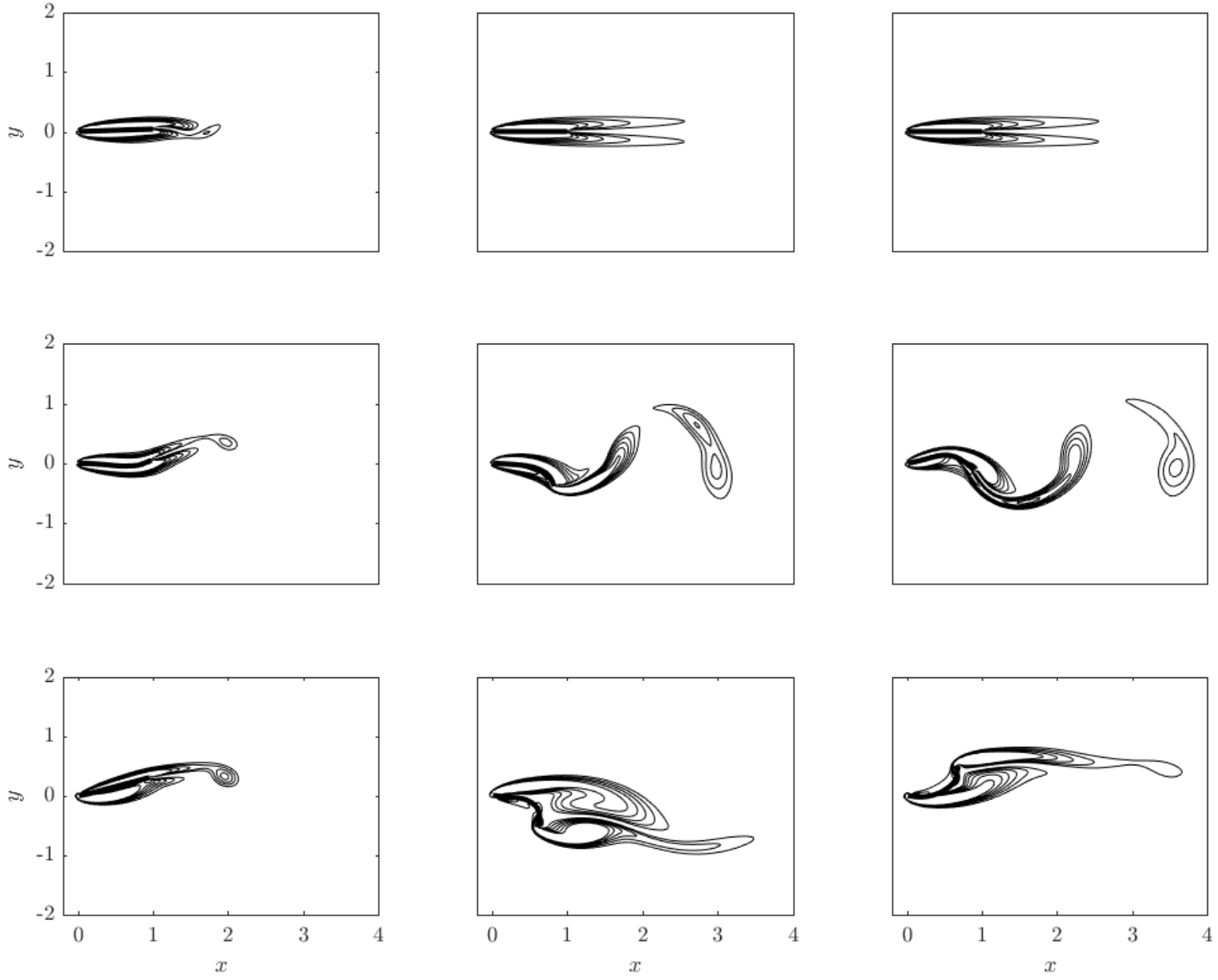}
    \caption{Vorticity contours of flow past a deforming beam for $\text{M}_\rho = 0.0001$ (top row), $\text{M}_\rho = 1.5$ (middle row), and $\text{M}_\rho = 100$ (bottom row). Left column: $t = 1.2$; middle column: $t = 18$; right column: $t = 31.2$. Contours are in 15 evenly spaced increments from $-5$ to $5$. The other parameters were $\text{Re} = 200$, $\text{M}_\rho = 1.5$, and $\text{Fr} = 1.4$.}
    \label{fig:Huang_LC_snap}
\end{figure}

In figure \ref{fig:Huang_LC_snap}, we show vorticity contours at different time instances for $\text{M}_\rho = 0.0001$, $\text{M}_\rho = 1.5$, and $\text{M}_\rho = 100$. When $\text{M}_\rho = 0.0001$, the impulsive start pushes the beam down quickly towards the $\theta = 0^\circ$ position. At $t\approx 0.5$, the vortical structure created during the impulsive start reaches the trailing edge of the beam. Due to the beam's small inertia, this substantially affects the trailing edge displacement (see insert in figure \ref{fig:Huang_LC_fig}). The vortical structure then advects away from the body, and the fluid wake becomes symmetric as the beam fully reaches its $\theta = 0^\circ$ position. This symmetric wake is a well known feature of flow past thin rigid bodies at low Reynolds numbers, and the small inertia of the beam does not allow for any minor beam deformations to break the flow symmetry. Thus, the beam stays in this neutral position for all remaining time. When $\text{M}_\rho = 1.5$, the limit cycle flapping of the beam is associated with a periodic vortex street \cite{huang,lee,eldredge15}. As the beam becomes increasingly massive, the flapping amplitude increases.  This is associated with a thicker, more irregular wake profile and more chaotic flapping behavior. Figure \ref{fig:Huang_LC_fig} shows that the $\text{M}_\rho = 100$ case does not enter into limit cycle flapping after 250 convective time units and several periods of flapping. The destabilizing nature of increasing the mass ratio was also noted by Connell \emph{et al.}\ \cite{connell}.

\subsection{Flow past a beam in the inverted configuration}

\begin{figure}[h!]
\centering
\includegraphics[width=1\textwidth]{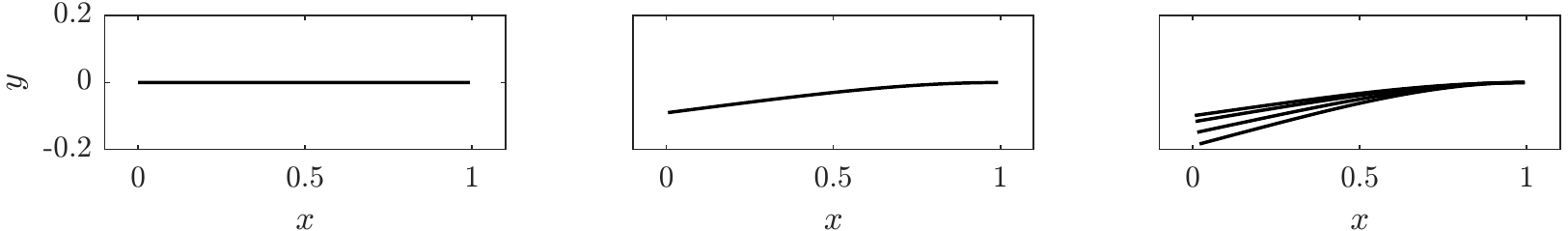}
\caption{Snapshots of the beam at various times for the fixed-point stable (left), static divergence (middle), and unstable flapping (right) regimes. Only one curve is visible for the fixed-point stable and static divergence cases, since the beam does not deflect about these equilibrium positions. }
\label{fig:gur_sch}
\end{figure}

For a beam initially placed at $\theta = 0^\circ$, there are three possible regimes: fixed-point stable, static divergence, or unstable flapping \cite{gurugubelli} (see figure \ref{fig:gur_sch} for an illustration). 

In table \ref{tab:gur_fla}, we summarize the regimes we obtained by varying $\text{K}_\text{B}$ for $\text{Re} = 200$, $\text{M}_\rho = 0.1$, and $\text{Fr} = 0$. Our finest sub-domain for this problem was of size $[-0.2,1.8]\times[-0.5,0.5]$, and the total domain size was $[-15.20,16.80]\times[-16,16]$. The grid spacing on the finest domain was $h = 0.002$, and the time step was $\Delta t = 0.00018$. Using $h = 0.0018$ change our reported results by less than one percent. Three iterations of (\ref{eqn:dx_fac}), (\ref{eqn:ffac_alt}), and (\ref{eqn:dufac_alt}) were required for the first time step, and a maximum of two iterations were required for all remaining time steps. The table also shows the regimes computed by Gurugubelli \emph{et al.}\ \cite{gurugubelli} for the same parameter ranges.
\begin{table}[h!]
\centering
\begin{tabular}{ c  !{\vrule width1.2pt}c  c  c  c  c  c  c  c  c  c  c  }
 & \multicolumn{11}{c}{  $\text{K}_\text{B}$ } \\ 
  & 0.41 & 0.42 & 0.43 & 0.44 & 0.45 & 0.46 & 0.47 & 0.48 & 0.49 & 0.50 & 0.51 \\ \Xhline{3\arrayrulewidth}
Gurugubelli \emph{et al.}\ \cite{gurugubelli} & UF & UF & UF & UF & UF & SD & SD & SD & SD & SD & FPS \\ 
Present & UF & UF & UF & SD & SD & SD & SD & SD & SD & SD & FPS \\ 
\end{tabular}
\caption{Flapping regimes obtained for different $\text{K}_\text{B}$ values. UF $=$ unstable flapping, SD $=$ static divergence, FPS $=$ fixed point stable. The other parameters were $\text{Re} = 200$, $\text{M}_\rho = 0.1$, and $\text{Fr} = 0$.}
\label{tab:gur_fla}
\end{table}

We next consider the case when $\text{Re} = 1000$, $\text{M}_\rho = 0.1$, $\text{K}_\text{B} = 0.4$, and $\text{Fr} = 0$. For this set of parameters, the beam enters large amplitude limit cycle flapping. Table \ref{tab:gur_LC} shows the amplitude and frequency of the leading edge transverse displacement computed in our work and in reference \cite{gurugubelli}. Our finest sub-domain for this problem was of size $[-0.2,1.8]\times[-1,1]$, and the total domain size was $[-15.20,16.80]\times[-14.35,14.35]$. The grid spacing on the finest domain was $h = 0.0033$, and the time step was $\Delta t = 0.0008$. Using $h = 0.0025$ did not change our reported results.
\begin{table}[h!]
\begin{center}
\begin{tabular}{ c !{\vrule width1.2pt}c   c  }
 & Amplitude & Frequency ($St$) \\ \Xhline{3\arrayrulewidth}
Gurugubelli \emph{et al.}\ \cite{gurugubelli} & $\pm 0.83$ & $0.204$ \\ 
Present & $\pm 0.82$ & $0.198$ \\
\end{tabular}
\caption{Amplitude and frequency associated with the transverse displacement of the beam's leading edge; obtained for Re $ = 1000$, M$_\rho = 0.5$, \text{K}$_\text{B} =0.4$, and Fr $ =0$.
}
\label{tab:gur_LC} 
\end{center}
\end{table}

\begin{figure}[h!]
\centering
\includegraphics[width=1\textwidth]{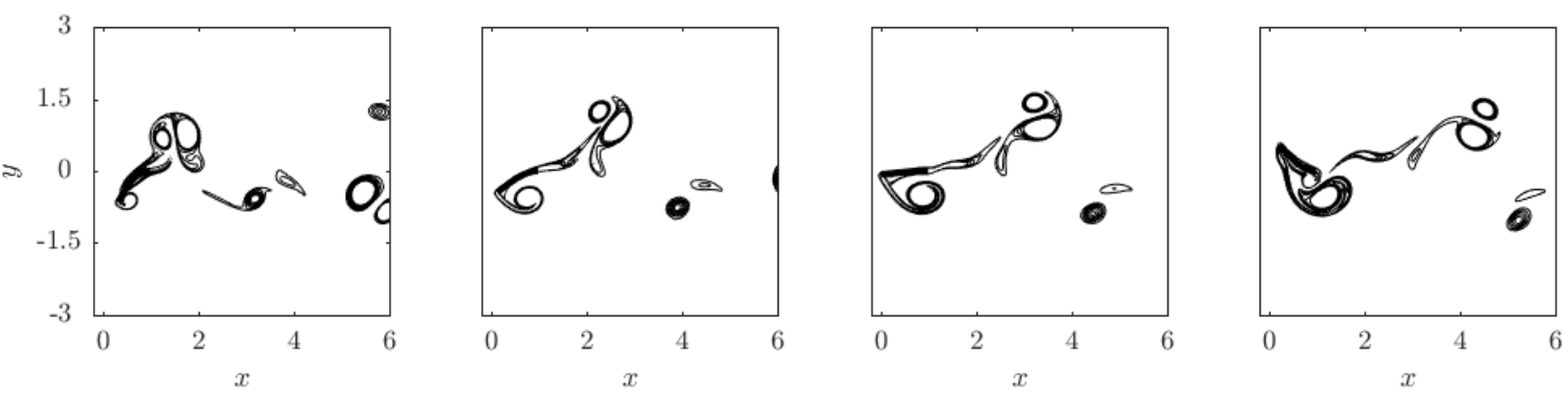}
\caption{Vorticity contours during a flapping cycle for $\text{Re}=1000$, $\text{M}_\rho = 0.5$, $\text{K}_\text{B} = 0.4$, $\text{Fr} = 0$. Contours are in 15 evenly spaced increments from $-5$ to $5$.}
\label{fig:gur_LC_fig}
\end{figure}

Figure \ref{fig:gur_LC_fig} shows vorticity contours at four different times during a flapping cycle. The figure shows features consistent with what was observed by Gurugubelli \emph{et al}\ \cite{gurugubelli}. The fluid stresses associated with the leading edge vortex deform the beam (leftmost figure). The beam begins to flap back towards the centerline once the internal beam stresses counteract the imposed fluid stresses and beam inertia, and the leading edge vortex detaches (second leftmost figure). The inertia of the beam causes it to continue to move towards the centerline while the detached vortex grows in size (second rightmost figure). The beam then flaps upward, and a vortex grows at the trailing edge (rightmost figure). After the snapshot in the rightmost figure, the detached leading and trailing edge vortices advect downstream and form a vortex pair of opposite sign to the one seen in the beam's near-wake in figure \ref{fig:gur_LC_fig}. The upward motion of the beam creates an attached leading edge vortex, which starts an analogous process to the one just described. The result of this upward-downward beam motion is limit cycle flapping that repeats indefinitely.

\section{Conclusions}

We presented an immersed-boundary method for fully coupled flow-structure interaction problems involving thin deforming surfaces. The method is strongly-coupled, and is therefore stable for wide ranges of solid-to-fluid mass ratios and large body motions. As with many strongly-coupled methods, our method requires the solution of a nonlinear system of equations at each time step. This system is solved by iteration, using a linearization of the nonlinearly coupled equations and a block LU factorization of the linearized system to reduce iterations to small-dimensional subsystems of equations. The iteration process does not involve heuristic relaxation parameters. We derived the method for general deforming surfaces, and verified it for 2D flow past deforming beams. The test problems involved a wide range of mass ratios and large body motions, and the method required a small number of iterations to converge for all cases considered. It is straightforward to extend the method to fully coupled problems involving rigid bodies.

\section{Acknowledgments}

This research was partially supported by a grant from the Jet Propulsion Laboratory (Grant No. 1492185). Many of the simulations were performed using the Extreme Science and Engineering Discovery Environment (XSEDE), which is supported by National Science Foundation grant number ACI-1053575. The first author gratefully acknowledges funding from the National Science Foundation Graduate Research Fellowship Program (Grant No. DGE-1144469). We thank Professor Jeff Eldredge for his helpful comments on the manuscript.

\begin{appendices}
\numberwithin{equation}{section}
\numberwithin{figure}{section}
\numberwithin{table}{section}

\section{An efficient iteration procedure in primitive variables}
\label{app:app}

Using primitive variables for the fluid, the differential-algebraic semidiscrete FSI system is
\begin{gather}
\dot{u} + \mathcal{N}(u) = -Gp + Lu -E^T(\chi) f \label{eqn:equ_app} \\
M \dot{\zeta} = -R(\chi) + Q(g + W(\chi)f) \label{eqn:eqzeta_app} \\
\dot{\chi} = \zeta \label{eqn:eqx_app} \\
Du = 0 \label{eqn:eqp_app} \\
E(\chi)u - \zeta = 0 \label{eqn:eqc_app} 
\end{gather}

Using the same time discretization schemes as in section 3 and introducing the decomposition $\chi_{n+1}^{(k+1)} = \chi_{n+1}^{(k)} + \Delta \chi$, $\zeta_{n+1}^{(k+1)}=\zeta_{n+1}^{(k)} + \Delta \zeta$, we have the following system to within first order in the increments and $\Delta t$:
\begin{equation}
\renewcommand\arraystretch{1.5}
\begin{bmatrix} A & G & 0 & 0 & E_{n+1}^{(k)\, T}\\ G^T & 0 & 0 & 0  & 0 \\ 0 & 0 & 0 & \hat{K}^{(k)} & -QW_{n+1}^{(k)} \\ 0 & 0 & -I & \frac{2}{\Delta t} I & 0 \\ E_{n+1}^{(k)} & 0 & -I & 0 & 0 \end{bmatrix} \begin{bmatrix} u \\ p_{n+1} \\ \Delta \zeta \\ \Delta \chi \\ f_{n+1}^{(k+1)} \end{bmatrix} = \begin{bmatrix} \tilde{r}^f_n \\ \tilde{r}^p_n \\ r^{\zeta \,(k)} \\ r^{\chi \,(k)} \\ r^{c\,(k)} \end{bmatrix}
\label{eqn:up_mat}
\end{equation}
where the right hand side terms are known and analogous to those in section 3. Performing a block-LU decomposition of (\ref{eqn:up_mat}) gives the sequence of equations
\begin{gather}
u^* = A^{-1}\tilde{r}^f_n - A^{-1} G(G^TA^{-1}G)^{-1}(G^T A^{-1}\tilde{r}^f_n - \tilde{r}^{p})\label{eqn:u*_fac} \\
p^* = (G^TA^{-1}G)^{-1}(G^TA^{-1} \tilde{r}^f_n - \tilde{r}^p_n ) \label{eqn:p*_fac} \\
\renewcommand\arraystretch{1.6}
\begin{bmatrix} \tilde{B}_{n+1}^{(k)} & I \\ -\frac{2}{\Delta t} QW_{n+1}^{(k)} & \hat{K}^{(k)} \end{bmatrix}  \begin{bmatrix}  f_{n+1}^{(k+1)} \\ \Delta \zeta \end{bmatrix} = \begin{bmatrix} E_{n+1}^{(k)} u^* - r^{c\,(k)} 
\renewcommand\arraystretch{2} \\ 
\frac{2}{\Delta t} r^{\zeta \,(k)} + r^{\chi \,(k)}\end{bmatrix} \label{eqn:fdu_fac2} \\ 
\Delta \chi = \frac{2}{\Delta t} (\Delta \zeta + r^{\chi \, (k)} ) \label{eqn:dx_fac2} \\
p_{n+1} = p^* -(G^TAG)^{-1}G^TA^{-1}E_{n+1}^{(k)\,T} f_{n+1} \label{eqn:p_fac} \\
u_{n+1} = u^* - A^{-1}(I - G(G^TA^{-1}G)^{-1}G^T A^{-1})E_{n+1}^{(k) \,T} f_{n+1} \label{eqn:u_fac} 
\end{gather}
where $\tilde{B}_{n+1}^{(k)} = E_{n+1}^{(k)}A^{-1}(I - G(G^TA^{-1}G)^{-1}G^TA^{-1})E_{n+1}^{(k)\,T}$.

As in section 3, all iterations are restricted to (\ref{eqn:fdu_fac2})--(\ref{eqn:dx_fac2}), which have dimensions on the order of the number of body points; (\ref{eqn:u*_fac})--(\ref{eqn:p*_fac}) may be computed once and for all at the start of a time step, and (\ref{eqn:p_fac})--(\ref{eqn:u_fac}) need only be solved after all iterations are completed to convergence.

Again, when $(\hat{K}^{(k)})^{-1}$ can be computed and stored, the system (\ref{eqn:fdu_fac2}) may be reformulated as the sequence of equations given by
\begin{gather}
\left(\tilde{B}_{n+1}^{(k)} + \frac{2}{\Delta t}(\hat{K}^{(k)})^{-1}QW_{n+1}^{(k)}\right)f_{n+1}^{(k+1)} = E_{n+1}^{(k)}u^{*} - r^{c\,(k)} -\frac{2}{\Delta t} (\hat{K}^{(k)})^{-1}  r^{\zeta\,(k)} + r^{\chi \,(k)}   \\
\Delta U = \frac{2}{\Delta t} (\hat{K}^{(k)})^{-1} ( r^{\zeta \,(k)} + QW_{n+1}^{(k)}f_{n+1}^{(k)}) - r^{\chi \,(k)}  
\end{gather}

\end{appendices}

\section*{References}

\bibliography{FSI_bib}

\end{document}